\documentclass[aps,prb,twocolumn,superscriptaddress,showpacs]{revtex4}

\usepackage[dvips]{graphicx}
\usepackage{color}
\usepackage{CJK}
\usepackage{bm}

\renewcommand{\vec}[1]{\boldsymbol{#1}}

\begin{document}
\begin{CJK*}{GB}{gbsn}

\title{Juxtaposition of Spin Freezing and Long Range Order in a Series of Geometrically Frustrated Antiferromagnetic Gadolinium Garnets}

\author{J.~A.~Quilliam} \altaffiliation{Present address: D\'{e}partement de Physique, Universit\'{e} de Sherbrooke, 2500, boul. de l'Universit\'{e}, Sherbrooke, QC J1K 2R1 Canada} 
\author{S.~Meng (ÃÏÊ÷³¬)} 

\affiliation{Department of Physics and Astronomy and Guelph-Waterloo Physics Institute, University of Waterloo, Waterloo, ON N2L 3G1 Canada}
\affiliation{Institute for Quantum Computing, University of Waterloo, Waterloo, ON N2L 3G1 Canada}

\author{H.~A.~Craig}\altaffiliation{Present address: Department of Applied Physics, Stanford University, Stanford, CA 94305 USA} 
\author{L.~R.~Corruccini}
\affiliation{Physics Department, University of California-Davis, Davis, California 95616, USA}

\author{G.~Balakrishnan}
\author{O.~A.~Petrenko}
\affiliation{Department of Physics, University of Warwick, Coventry CV4 7AL, United Kingdom}

\author{A.~Gomez}
\author{S.~W.~Kycia}
\affiliation{Department of Physics and Guelph-Waterloo Physics Institute, University of Guelph, Guelph, ON N1G 2W1, Canada}

\author{M.~J.~P.~Gingras}
\affiliation{Department of Physics and Astronomy and Guelph-Waterloo Physics Institute, University of Waterloo, Waterloo, ON N2L 3G1 Canada}
\affiliation{Canadian Institute for Advanced Research, 180 Dundas Street West, Suite 1400 Toronto, ON M5G 1Z8 Canada}

\author{J.~B.~Kycia}
\affiliation{Department of Physics and Astronomy and Guelph-Waterloo Physics Institute, University of
Waterloo, Waterloo, ON N2L 3G1 Canada}
\affiliation{Institute for Quantum Computing, University of
Waterloo, Waterloo, ON N2L 3G1 Canada}

\date{\today}

\begin{abstract}

Specific heat measurements in zero magnetic field are presented on a homologous series of geometrically frustrated, antiferromagnetic, Heisenberg garnet systems. Measurements of Gd$_3$Ga$_5$O$_{12}$, grown with isotopically pure Gd, agree well with previous results on samples with naturally abundant Gd, showing no ordering features.  In contrast, samples of Gd$_3$Te$_2$Li$_3$O$_{12}$ and Gd$_3$Al$_5$O$_{12}$ are found to exhibit clear ordering transitions at 243 mK and 175~mK respectively.  The effects of low level disorder are studied through dilution of Gd$^{3+}$ with non-magnetic Y$^{3+}$ in Gd$_3$Te$_2$Li$_3$O$_{12}$.  A thorough structural characterization, using X-ray diffraction, is performed on all of the samples studied.  We discuss possible explanations for such diverse behavior in very similar systems.

\end{abstract}

\pacs{75.50.Lk, 75.50.Ee, 75.40.Cx}
\keywords{}

\maketitle
\end{CJK*}


\section{Introduction}

Magnetic systems of spins residing on the sites of a lattice of corner-sharing simplexes, such as triangles or tetrahedra, coupled via a nearest-neighbor (n.n.) antiferromagnetic Heisenberg exchange Hamiltonian, $\mathcal{H}_0$, are highly frustrated.  The spins are unable to minimize their energy pair by pair and, at the classical level, such a model tends to result in a sort of ``spin liquid'' state without long range order (LRO) and with zero net magnetic moment, ${\bf M}_s$, on each simplex.~\cite{Moessner2001} Perturbations to $\mathcal{H}_0$, $\mathcal{H}'$, such as exchange beyond n.n. and dipolar interactions, perhaps assisted by thermal and/or quantum fluctuations, are typically expected to lift the classical ground state degeneracy and drive the system into a state of LRO.  However, the ground state remains extremely fragile against quenched random disorder and as a result, rather than developing LRO, the combination of high frustration and weak random disorder can in principle~\cite{Villain1979,Andreanov2010,BellierCastella2001} cause a system to exhibit a spin glass transition.  Such process seems to be occurring in the pyrochlore system Y$_2$Mo$_2$O$_7$ (YMO),~\cite{Gingras1997,Keren2001} for example.

\begin{figure}
\begin{center}
\includegraphics[width=2.75in,keepaspectratio=true]{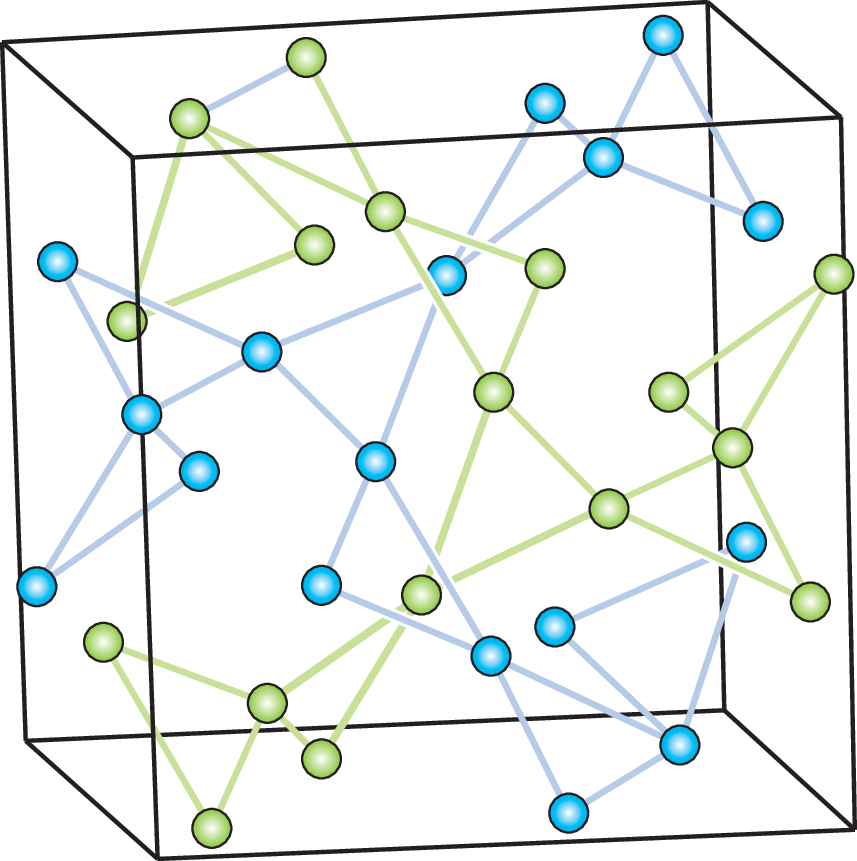}
\caption{(color online) The cubic unit cell of the Gd$_3$A$_2$B$_3$O$_{12}$ garnet structure, showing only the Gd positions which form two interpenetrating networks of corner-sharing triangles, known as hyperkagome lattices (one shown in blue and the other in green). }
\end{center}
\end{figure}

In this article, we investigate a homologous series of frustrated garnet materials which are described by a 3-dimensional network of corner sharing triangles~\cite{Hyperkagome} (shown in Fig. 1) which, despite their  similar Hamiltonians, exhibit very diverse thermodynamic behavior.  Specifically, we carry out low temperature  specific heat measurements on three Gd garnet materials: Gd$_3$Ga$_5$O$_{12}$ (GGG), Gd$_3$Te$_2$Li$_3$O$_{12}$ (GTLG) and Gd$_3$Al$_5$O$_{12}$ (GAG).  With largely isotropic, $S=7/2$ Gd$^{3+}$ moments, these materials should be fair representations of classical Heisenberg antiferromagnets.  However, because of the low energy scale of exchange interactions ($J_1 \sim 0.1$ to 0.14 K),\cite{Applegate2007} the dipolar interaction becomes very important.  

\begin{figure}
\begin{center}
\includegraphics[width=3.25in,keepaspectratio=true]{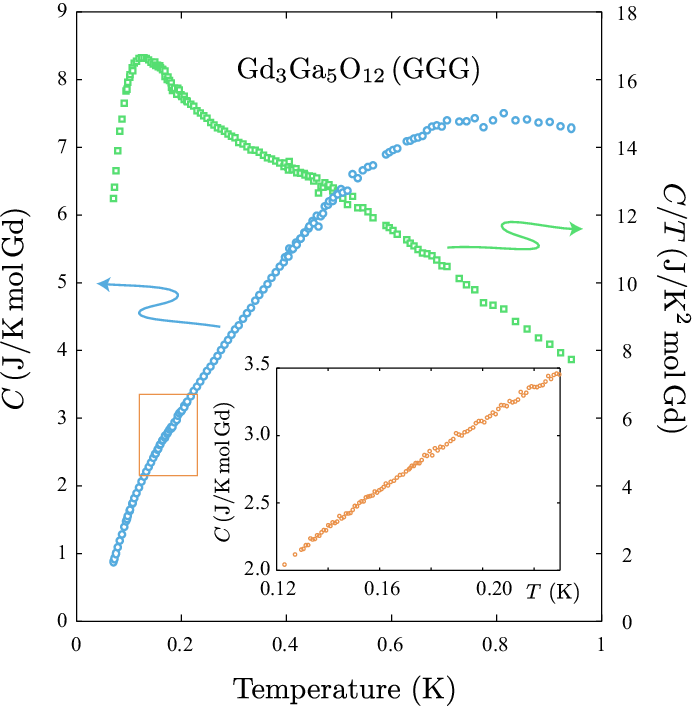}
\caption{(color online) Specific heat ($C$) and $C/T$ of isotopically pure GGG measured with coarse temperature resolution.  Results match well with previous work~\cite{Schiffer1995}.  The temperature region marked with a red box has been remeasured with a high temperature resolution shown in the inset and again does not find evidence of an ordering transition. \label{gggFigure} }
\end{center}
\end{figure}

The first material, GGG, has been extensively studied previously.  Experiments by Schiffer \emph{et al.} have suggested that GGG is a spin glass in low field, albeit a somewhat unconventional one.~\cite{Schiffer1995}  The spin glass interpretation is largely based on glassy relaxation in the ac susceptibility and a sharp peak in the nonlinear susceptibility ($\chi^3$) at around 180 mK.~\cite{Schiffer1995}  However, there is also a broader peak in $\chi^3$ at 450 mK and there is no corresponding maximum in the specific heat (as seen in Fig.~\ref{gggFigure}) but rather a maximum in $C/T$ at around 125 mK.  Applying small magnetic fields, $\gtrsim 300$ mT, destroys the spin glass state and gives way to a cooperative paramagnetic or spin liquid state.~\cite{Schiffer1994}  At higher magnetic fields, $\gtrsim 0.6$ T, a complex phase diagram emerges including a ``bubble'' of antiferromagnetic order and reentrant behavior reminiscent of the $^4$He melting curve.~\cite{Tsui1999,Tsui2001CJP}

One might quickly conclude that the unconventional glassy physics of GGG is a result of small levels of disorder.  Indeed GGG is known to have an unavoidable 1-2\% off-stoichiometry whereby excess Gd is found on some of the Ga sites.~\cite{Daudin1982}  However, the glassy behavior of YMO, for instance, is fairly standard~\cite{Gingras1997} whereas GGG shows highly unconventional properties suggesting that the glassiness of GGG has a more complicated origin than a simple sensitivity to disorder.  Therefore, in this work, we aim to explore the effects of subtle changes in Hamiltonian and disorder on the Gd-garnet lattice. To do so, we have measured the specific heat of a series of homologous Gd garnet materials, all of which are highly frustrated, yet the resulting ground states are found to be very different, with clear indications of long range order (LRO) in all samples except GGG.  Hence, we show that the exotic physics of GGG is not an inherent property of Gd garnets and results from much more subtle effects.  Additionally, we have tested the effects of random chemical substitution of non-magnetic Y$^{3+}$ for Gd$^{3+}$ in GTLG to investigate in a controlled manner a level of disorder comparable to, or even higher than, that of GGG.  Our results show that spin glass behavior resulting from a generic sensitivity to an arbitrary form of disorder is too simple a picture for this particular series of materials.  

In this paper, we first discuss the samples studied, including their likely Hamiltonians, and summarize a detailed structural characterization which is further detailed in the Appendix.  We present results on a GGG sample which support previous work and then discuss new results on GTLG and GAG as well as the effects of controlled substitutional disorder.  Finally, we discuss possible explanations for the diverse thermodynamic behavior seen in this very similar set of materials.

\section{Samples and Experiment}

We have measured a single crystal of GGG, made from isotopically pure $^{160}$Gd, grown from the same powder used for neutron scattering measurements.~\cite{Petrenko1997, Petrenko1998}  Preparation of this sample has been described in Ref.~\onlinecite{Petrenko2009}.  The samples of GTLG and GAG were polycrystalline samples made from Gd with natural isotopic abundance.  GTLG samples were made by solid-state reaction in air of Te and Gd (Y) oxides and Li carbonate, pressed into pellets, at 850 C for 10 hours, then regrinding and firing a second time.  The polycrystalline sample of GAG was made using the sol-gel method.~\cite{Costa2007}  It was heated to 1350 C for one hour, then removed from the furnace and rapidly quenched to room temperature, to minimize the formation of perovskite-phase GdAlO$_3$.~\cite{Chiang2007}  Powder x-ray diffraction spectra in both cases could be indexed to space group $\mathrm{Ia\bar{3}d}$.  A 2.4\% (by weight) GdAlO$_3$ impurity phase was found in GAG.

GGG has been the subject of numerous measurements in the past as a result of its exotic low-temperature behavior. It is well understood that GGG samples tend to suffer from an off-stoichiometry, specifically excess Gd$^{3+}$ ions randomly occupying otherwise non-magnetic Ga$^{3+}$ sites.\cite{Darby2008,Allibert1974,Daudin1982,Schiffer1994}  The level of this disorder is typically 1 to 2\% (see Ref.~\onlinecite{Daudin1982}) thus the chemical formula of GGG is more accurately written as Gd$_{3+x}$Ga$_{5-x}$O$_{12}$, with $x \simeq 0.03$ to 0.06.  In one attempt to control the off-stoichiometry of GGG, Schiffer \emph{et al.}~\cite{Schiffer1995} grew a powder sample rather than a single crystal, in the hope that the sample's stoichiometry would better match that of the starting ingredients (which was correct to within 0.1\%).  Although this resulted in a 50 mK reduction in the freezing temperature, it is not clear whether the change is a result of a modified stoichiometry or some other structural parameter.  Specific heat measurements by Dunsiger \emph{et al.}~\cite{Dunsiger2000GGG} on a powder sample of GGG gave essentially the same results as single crystal experiments.  Without being able to appreciably alter the inherent disorder in GGG, we propose here to explore the results of subtle changes in the Gd-based garnet Hamiltonian by studying the homologous materials GTLG and GAG, that may also be cleaner systems.

We have engaged in an in-depth characterization of all of the samples measured here in an attempt to quantify the disorder in GGG and to determine whether the other garnet systems (GTLG and GAG) are in fact ``cleaner'', with regard to substitutional disorder.  At least naively, it does seem improbable for Gd$^{3+}$ ions to easily substitute for Li$^{+}$ or Te$^{6+}$ ions in GTLG, which possess very different valence charges.  In the case of GAG, the valence of Gd$^{3+}$ and Al$^{3+}$ ions is the same, but the ionic radius of Al$^{3+}$ is quite different, more so than Ga$^{3+}$ for example.\cite{Crystallography,LanthAct}  It remains important to validate such assumptions with a structural characterization of the materials.  A complete description of the X-ray diffraction study is given in the Appendix, but we include here a summary of the salient results.

In the case of single crystal GGG, no diffuse scattering was detected, which would have been an indication of chemical disorder or stacking faults in the sample.  High resolution reciprocal space mapping around Bragg reflections showed very clean resolution-limited symmetric peaks with no appreciable effects due to off-stoichiometry domains in the sample.  If such domains exist, they must be larger than $\sim 150$ nm in size and exhibit lattice constants identical to the rest of the GGG sample.  X-ray diffraction measurements are thus consistent with the GGG crystal being highly perfect.   Given that there is likely some off-stoichiometry in our GGG sample, the contrast between Gd and Ga is perhaps not sufficient to permit the direct observation of such disorder in the crystal structure.  However, an important piece of information comes from the lattice constant, determined for our sample to be $12.3873 \pm 0.0001$ \AA at room temperature.  In previous structural work on GGG, a linear correlation between excess Gd concentration, $x$, in the formula Gd$_{3+x}$Ga$_{5-x}$O$_{12}$, and the lattice parameter $a$ was observed.\cite{Darby2008,Allibert1974}  Increasing Gd concentration contributes to a systematic and noticeable increase in the lattice parameter allowing us to indirectly determine the stoichiometry of our sample.  Based on that work, we obtain $x = 0.053 \pm 0.005$ for our sample, or $1.8\% \pm 0.1\%$ excess Gd$^{3+}$ on the Ga sites in our GGG crystal.

Powder X-ray diffraction patterns of the polycrystalline GTLG and GAG samples also exhibit very sharp peaks and Rietveld refinements showed excellent agreement with the expected crystal structure. Pure and Y-doped GTLG samples showed very similar lattice constants: 12.3865 (5) \AA\, and 12.3861(5) \AA, respectively.  This suggests that the Y$^{3+}$ ions easily take the place of Gd$^{3+}$ ions, without otherwise perturbing the crystal structure. A Rietveld fit showed the Y-content in the Y-doped sample to be 3.2(9)\%, close to the expected target concentration of 2\%.  The lattice parameter of our GAG sample is 12.1090(6) \AA.

While no indications of an off-stoichiometry were observed, the error in the fit gives a maximum possible level of off-stoichiometry of 0.4\% on the Al (in GAG) or Li (in GTLG) sites.
Thus, without being able to completely rule out an excess of Gd on Al or Li sites (in GAG and GTLG respectively), we can state that it is \emph{at most} a fairly small level compared to the off-stoichiometry in GGG.  However, there is insufficient contrast in the X-ray measurements between Gd and Te (in GTLG), to determine whether there is mixing between those sites.  

As a good method of discerning long range order (LRO) from conventional spin glass physics (in which one finds broad specific heat bumps~\cite{Binder1986}) and moreover from the unconventional glassy behavior previously seen in GGG, we used specific heat, $C$, measurements to characterize the magnetic ground states of these systems.  We employ the quasi-adiabatic method as described in Refs.~\onlinecite{Quilliam2007,Quilliam2007GSO}, with thermometer and heater fixed directly to the samples.  6 $\mu$m diameter, $\sim 1$ cm long superconducting leads were used to provide excellent thermal isolation.  A long time constant of relaxation ($\tau > 1$ hour) ensured that internal temperature gradients due to poor thermal conductivity (of particular concern in the powder samples) were not a significant source of systematic error.

\section{Hamiltonians}

Before comparing the properties of these materials, it is important to discuss how similar one expects them to be at the microscopic level.  They are likely to be described primarily by a Hamiltonian consisting of nearest-neighbor exchange 
\begin{equation}
\mathcal{H}_\mathrm{Ex} =  \sum_{\langle i,j\rangle} J_1 \vec{S}_i \cdot \vec{S}_j
\end{equation}
and dipolar interaction
\begin{equation}
\mathcal{H}_D =  D \sum_{\langle i,j \rangle} \frac{r_{nn}^3}{r_{ij}^3} \left[ \vec{S}_i\cdot\vec{S}_j - 3(\vec{S}_i\cdot \hat{\vec{r}}_{ij})(\vec{S}_j \cdot \hat{\vec{r}}_{ij}) \right]
\end{equation}
Further neighbor interactions may be important, as seems to be the case in GGG,\cite{Yavorskii2006,Yavorskii2007} but for the other compounds (GTLG and GAG) studied here, these interactions have not yet been determined.  There also may be some small single-ion anisotropy present in these materials.  The single-ion anisotropy energy of GGG was found to be less than 0.04 K~\cite{Overmeyer}  and it might be different in the other garnets, especially in GTLG which has a different charge configuration hence a possibly different crystal field.    

Measurement of the high temperature susceptibility gives the Curie-Weiss temperature $\theta_{CW}$, which is insensitive to the dipolar interaction for an isotropic moment,~\cite{Applegate2007} but instead represents primarily the nearest-neighbor exchange interaction, obtained through $\theta_{CW}   \simeq 4J_1nS(S+1)/3k_B$, where $n=4$ is the coordination number.\cite{Applegate2007}  This gives  $J_1 = 126$ mK for GTLG,~\cite{Applegate2007} $J_1 = 142$ mK for GAG,~\cite{Wolf1962} and $J_1= 107$ mK for GGG.~\cite{Kinney1979,Yavorskii2006}  The dipolar interaction, on the other hand, is purely defined by the size of the moments and the distance between the magnetic ions.  Since the lattice parameters of these compounds are very similar and we have the same magnetic species in each material, the dipolar interactions are also very similar in magnitude.  In fact, it is nearly identical between GGG and GTLG at $D = 45$ mK whereas  GAG has a slightly increased dipolar interaction of $D=48$ mK.  Thus, across this series of materials we have a subtle progression of the ratio $D/J_1$: from GAG (0.34) to GTLG (0.36) then GGG (0.42).  The known Hamiltonian parameters of these systems are summarized in Table~\ref{HamParams}.

\begin{table}
\caption{Nearest-neighbour exchange interaction ($J_1$), lattice parameter ($a$), dipolar interaction strength ($D$) and transition temperature ($T_C$) for the four materials studied here.
\label{HamParams}
}
\begin{ruledtabular}

\begin{tabular}{lllll}
Compound   & $J_1$ (mK) & $a$ (\AA)  & $D$ (mK)  & $T_C$ (mK) \\
\hline

Gd$_3$Ga$_5$O$_{12}$  	    & 107       & 12.387    & 45       &   140 ($T_g$)\\
Gd$_3$Li$_2$Te$_3$O$_{12}$ & 126      & 12.387     & 45      &  243 \\
Gd$_3$Al$_5$O$_{12}$ 	 	    & 142      & 12.109     & 48      &  175  \\

\end{tabular}

\end{ruledtabular}
\end{table}

\section{Results}

\textbf{\textit{GGG Results.}} An initial measurement of $C(T)$ of GGG was made over a large temperature range (from 80 mK to 930 mK) using temperature steps of roughly 5 mK below 200 mK and temperature steps of 10 mK above 200 mK.  Results of our specific heat measurements on GGG, shown in Fig.~\ref{gggFigure}, agree remarkably well with the previous specific heat measurement of Schiffer \emph{et al.}~\cite{Schiffer1995} (on a single crystal containing naturally abundant Gd).  We find a broad feature with a maximum at around 800 mK.  This feature seems to drop out at lower temperature roughly as $T^{0.8}$ until around 125 mK, at which point there is a maximum in $C/T$ as the specific heat develops a steeper $T$-dependence.  The specific heat measurement of Dunsiger \emph{et al.} on a naturally abundant Gd, powder sample is very similar, though the peak in $C/T$ is found to be more pronounced.~\cite{Dunsiger2000GGG}  This result suggests that the isotopically pure sample measured here exhibits the same physics as do naturally abundant Gd-containing samples and reconfirms the absence of a sharp ordering feature in $C$ that would indicate a transition to LRO.

While most early low $T$ experiments on GGG show spin glass-like behavior, there are a number of later results that provide evidence \emph{against} a conventional spin glass transition.  Muon spin relaxation ($\mu$SR) and M\"{o}ssbauer spectroscopy experiments, for example, show significant persistent spin dynamics (PSDs) down to temperatures as low as 25 mK.\cite{Dunsiger2000GGG,Marshall2002,Bonville2004GGG}  Most curiously, neutron scattering experiments on GGG, made possible by samples made from isotopically pure $^{160}$Gd that does not absorb neutrons, have shown sharp diffraction peaks developing below 140 mK.~\cite{Petrenko1997, Petrenko1998}.   Though these peaks are not sharp enough to imply true long range order, they suggest magnetic ordering with a correlation length of at least 100~\AA.  This has been suggested to be a type of mixed spin liquid/solid state -- a ``spin slush'' of sorts.~\cite{Petrenko1998}  More recent bulk ac susceptibility measurements have shown highly unconventional glassy relaxation and have been interpreted as a signature of an ordering transition.~\cite{Ghosh2008}  Recent inelastic neutron scattering measurements have revealed three gapped dispersionless excitations, two of which have been attributed to dimerized antiferromagnetic correlations.\cite{Deen2010}

In a theoretical study,~\cite{Yavorskii2006,Yavorskii2007} aiming to relate the sharp features in the GGG neutron scattering pattern~\cite{Petrenko1998} with a long-range ordered state, the spin-spin correlations were treated via a mean-field theory and using an Ewald summation method to handle the important dipolar interactions.  Tuning the second ($J_2$) and third ($J_3$) n.n. couplings and simulating the powder neutron diffraction signal, Ref.~\onlinecite{Yavorskii2006} was able to find excellent agreement with experiment, thereby determining optimal interaction strengths.  Most interestingly, it was found that the system exhibits a quasi-degeneracy critical (soft) modes that would lead to an enhancement of thermal and quantum fluctuations as well as make the system very sensitive to small amounts of disorder.~\cite{Yavorskii2007,Villain1979}


This apparent development of rather long range order~\cite{Petrenko1998} and its seeming consistency with theoretical expectation~\cite{Yavorskii2006,Yavorskii2007} makes the lack of an ordering transition in specific heat measurements very paradoxical indeed.  In order to address this problem, we performed an additional measurement using a much higher temperature resolution of 1 mK over the temperature range 130 mK to 230 mK to search for small or narrow features near where the neutron scattering peaks were discovered that might have been previously missed.  This choice of resolution is based on scaling results of well characterized antiferromagnets to a transition temperature of 140 mK, suggesting one might expect a peak in $C$ with a width of only several mK.  However, the high-resolution scan of $C$, shown in the inset of Fig.~\ref{gggFigure}, also does not reveal any anomalies that might be interpreted as an ordering transition.

\begin{figure}
\begin{center}
\includegraphics[width=3.25in,keepaspectratio=true]{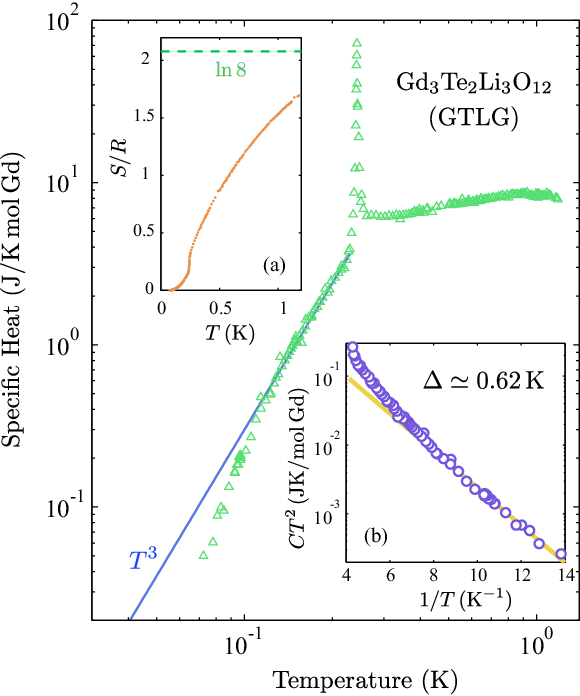}
\caption{(color online) Specific heat of GTLG showing a sharp first-order phase transition.  The specific heat drops out quicker than $T^3$ below $T_c$.  Inset (a) shows the entropy as a function of $T$ compared with the total $R\ln8$ entropy in the system.  Inset (b) shows $CT^2$ as a function of $T^{-1}$ and a linear fit suggestive of spin waves with a gap of $\Delta = 0.62$~K. \label{GTLGFigure} }
\end{center}
\end{figure}

\textbf{\textit{GTLG Results.}} Having again verified that GGG lacks an ordering transition, we turn to very similar systems that may shed light on its unusual behavior.  Thus we perform measurements on the related materials GTLG and GAG.  In stark contrast to GGG, GTLG displays a very sharp phase transition at 243 mK as shown in Fig.~\ref{GTLGFigure}.  This is close to the temperature ($\sim 250$ mK) where a feature was previously observed in the magnetic susceptibility of GTLG.~\cite{Applegate2007}  The transition is clearly first-order, exhibiting a much sharper peak in $C$ than could be expected for a continuous phase transition.  With a Curie-Weiss temperature $\theta_{CW}\simeq - 2.7$ K in GTLG,~\cite{Applegate2007} this ordering temperature gives a frustration index $f = \theta_{CW}/T_C = 11$.   Below the transition, $C$ drops out faster than $T^3$ suggesting that it is exponential and that there are gapped spin wave excitations, as would be expected from LRO with a strong dipolar interaction and as previously reported in the frustrated pyrochlore antiferromagnet Gd$_2$Sn$_2$O$_7$.~\cite{Maestro2007,Quilliam2007GSO}  Gapped spin waves should result in a low temperature behavior $C \sim T^{-2}e^{-\Delta/T}$, thus we plot $\log(CT^2)$ vs. $1/T$ in inset (b) of Fig.~\ref{GTLGFigure}.  The resulting linear fit gives $\Delta \simeq 0.62$ K.  As in Gd$_2$Sn$_2$O$_7$, we expect the gap here to come primarily from the anisotropic dipole-dipole interaction $\mathcal{H}_D$. Above the transition, there is a broad feature centered around roughly 1.0~K, similar to the broad feature in GGG centered at $\sim 0.8$ K (see Fig.~\ref{gggFigure}).  This broad feature is likely related to the gradually developing short range correlations which are observed for $T\lesssim 3$ K with neutron diffraction experiments in the case of GGG.~\cite{Petrenko1998}  Obtaining the entropy ($S$) from a  numerical integral of $C/T$ shows that only about 14\% of the total $R\ln8$ entropy in the system is accounted for by the transition, as shown in inset (a) of Fig.~\ref{GTLGFigure}. 

\begin{figure}
\begin{center}
\includegraphics[width=3.25in,keepaspectratio=true]{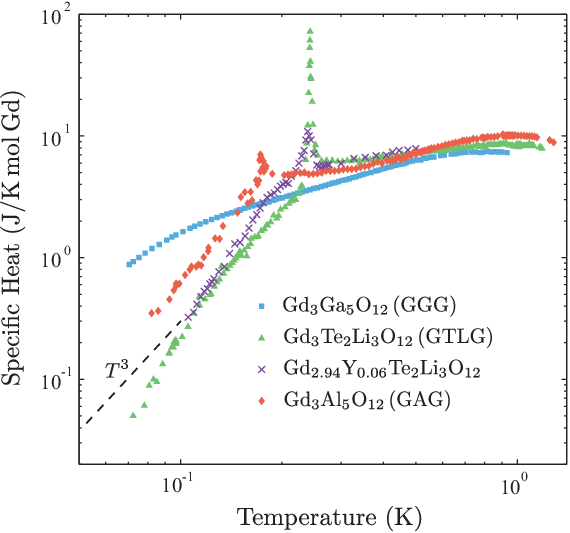}
\caption{(color online) Comparison of the specific heat of GGG (blue squares), GTLG (green triangles), GAG (red diamonds) and a sample of 2\% diluted GTLG (violet x's). \label{comparisonFigure}}
\end{center}
\end{figure}

\textbf{\textit{GAG Results.}} The third system studied here, GAG, may represent a ``material bridge'' in between GGG and GTLG with a smaller and broader transition at a lower temperature of 175 mK (shown in Fig.~\ref{comparisonFigure}).  With $\theta_{CW} \simeq -3.0$ K,~\cite{Wolf1962} it is more antiferromagnetic than the other two garnets, but seemingly more frustrated than GTLG, with a frustration index $f = 17$.  Otherwise, it shows similar features to GTLG, with a broad maximum centered around 1 K and a steeply dropping specific heat at lower temperatures.  As in GTLG, the transition in GAG accounts for only a small percentage of the total $R\ln 8$ entropy in the system.  The smaller and broader peak may result from the small ($\sim 3\%$) fraction of GdAlO$_3$ impurity phase mentioned in Section II.

\textbf{\textit{Effect of Dilution.}} It seems likely that GTLG does not have the same off-stoichiometry that is found in GGG, making it a ``cleaner system''; Gd$^{3+}$ ions are unlikely to take the place of Li$^+$ or Te$^{6+}$ ions as easily as Ga$^{3+}$ ions and the extensive sample characterization discussed above and in the Appendix corroborate such an expectation.  In order to explore the effects randomness, we have added a low level of disorder to GTLG through dilution of Gd$^{3+}$ ions with 2\% non-magnetic Y$^{3+}$ ions.  The level of disorder introduced in this way is comparable to the off-stoichiometry measured in our sample of GGG, and which is typical of GGG samples.  Nonetheless, this ``dirty'' GTLG sample shows a significantly broadened peak but \emph{no} noticeable change in the peak temperature, as shown in Fig.~\ref{comparisonFigure}.  At temperatures well below and well above the transition, the specific heat of the pure and diluted samples match, and the transition region (from around 140 to 300 mK)  accounts for the same amount of entropy in both systems, suggesting that the ground state ordering is not appreciably altered by the 2\% impurity. 

\section{Discussion}

Despite the expectation that they are described by similar microscopic spin Hamiltonians, we find GTLG and GAG to in fact behave entirely differently from GGG.  The only commonality in all three systems is the broad feature signaling short range correlations at around 0.8 K in GGG and 1.0 K in GAG and GTLG.  The lower temperature of the broad feature in GGG is consistent with its smaller nearest neighbor exchange interaction $J_1=0.107$ K~\cite{Kinney1979,Yavorskii2006} as compared to $J_1 = 0.126$ K for GTLG~\cite{Applegate2007} and $J_1 = 0.142$ K for GAG.~\cite{Wolf1962}

The sharp features observed in GTLG and GAG are in all likelihood signatures of transitions to long range order.  In fact the majority of insulating rare-earth garnets studied~\cite{Onn1967,Landau1971,Kamazawa2008} exhibit transitions to a magnetically ordered state.~\cite{Note2}  An exponential drop in the specific heat of GTLG below the transition, indicative of  gapped spin-wave excitations, also provides strong evidence of LRO.  Such behavior is reminiscent of the Gd pyrochlore material Gd$_2$Sn$_2$O$_7$ which also shows a sharp first-order phase transition to a ground state exhibiting static magnetic order~\cite{Maestro2007} and well-defined, gapped excitations as seen by neutron scattering~\cite{Wills2006} and specific heat experiments.~\cite{Quilliam2007GSO}  

The materials GTLG and GAG clearly do not share the same glassy physics as GGG since spin glasses are universally found \emph{not} to exhibit a sharp peak in $C$, but rather a broad feature near $T_g$.~\cite{Binder1986}  Our results prove that the glassy physics of GGG is not a ubiquitous property of Gd garnets and we seem therefore to be left with two possible conclusions.  

(i)  The first possibility is that GGG exhibits a spin glass transition that is a result of its finely tuned Hamiltonian and that is unrelated to the small levels of quenched disorder present in the system.  The theoretical support for such a topological spin glass state is severely limited, however, one example being anisotropic kagome antiferromagnets.~\cite{Note}

Nonetheless, it is tempting to consider an exciting possibility: that the presence of several interaction terms of different spatial range and anisotropic nature may, through their competition, lead to a complex energy landscape causing glassy behavior that is \emph{not} induced by quenched randomness.~\cite{Note}  Such a phenomenon parallels a theoretical description of \emph{structural} glasses that builds on the notion of locally preferred structure in a liquid that is frustrated at large length scale due to its inability to suitably fill space, hence inhibiting the formation of a crystalline state and leading to a glass transition.~\cite{Tarjus2005}  Could a HFM system, with its locally satisfied ${\bf M}_s=0$ simplexes, display a disorder-free freezing akin to the glass transition when subject to the competition of the longer-range terms that constitute $\mathcal{H}'$?  This may well be realized in a system like GGG where the dominant local, nearest neighbor Heisenberg exchange (constituting $\mathcal{H}_0$) is in competition with the also important long range dipolar interaction, representing $\mathcal{H}'$.~\cite{Yavorskii2006,Yavorskii2007}  If such a scenario is applicable, the relatively sharp neutron diffraction peaks~\cite{Petrenko1998} may constitute a signature of the short range locally preferred structure.  

Within such a picture, some accidental fine tuning of the parameters of the Hamiltonian is likely required in order to produce the necessary competition between local and long range interactions to give rise to a magnetic analog of the glass transition, even without quenched disorder.  GTLG and GAG may be outside this narrow ``window'' of required parameter space, with too low a ratio of $D/J_1$ to create the required long-range frustration.  However, in many glass forming liquids (see for example Ref.~\onlinecite{Alba1999}) it is found that slightly tuning interaction strengths, for instance by altering the substituents of molecules, without changing the overall symmetry of those molecules, does not tend to preclude the glass transition.  The two ordered materials, GTLG and GAG, show a more conventional behavior of frustrated dipolar antiferromagnets, where the dipolar interaction \emph{relieves} frustration and increasing $D/J_1$ leads to increasing $T_C/J_1$.  However, limited conclusions can be drawn from the ratio $D/J_1$ alone, since further-neighbor exchange interactions are likely to play an important role in the physics of Gd-based garnets.  Indeed, it was found in mean-field calculations that the paramagnetic correlations in a Heisenberg spin garnet system are highly sensitive to second and third nearest neighbor exchange.\cite{Yavorskii2006}


(ii) A plausible alternative is that a sensitivity of the system to small levels of disorder indeed results in a spin glass transition in GGG.  In this context, since the ordered ground state of GTLG appears to be robust against a level of random magnetic site dilution at least as high as that found in GGG, our results perhaps indicate that the \emph{type} of disorder may be the crucial ingredient causing GGG's exotic behavior.  In other words, the random excess of Gd on Ga sites, found in GGG, may be a much more powerful way to introduce random frustration and trigger a spin glass transition than simple dilution of the magnetic moments.  

In conclusions, we have provided some evidence, on the basis of X-ray diffraction measurements, that this off-stoichiometry is not present or is at least less significant in the GTLG and GAG systems, which instead show clear ordering transitions.  This work may, therefore, suggest that the response of a HFM to weak random disorder can be highly nontrivial, and a function of the nature of that disorder in concurrence with the nature of the interactions present.

\begin{acknowledgments}
Thanks to T.~Yavors'kii, G.~Tarjus and P.~H.~Poole for very informative conversations and to C.~G.~A.~Mugford for laboratory support.  Funding was provided by NSERC, CRC, CFI, MMO and Research Corporation grants as well as the Canada Research Chair program (M. J. P. G., Tier 1).
\end{acknowledgments}

\appendix*
\section{Details of Sample Characterization}

\subsection{Powder samples}

The powder samples were finely ground in an agate mortar and pressed into an aluminium sample holder. Powder X-ray diffraction patterns (XRD) were collected in Bragg-Brentano geometry at room temperature in a STOE goniometer using the Cu K$\alpha$ radiation ($\lambda$=1.54178 \AA) produced by en ENRAF-NONIUS FR571 rotating anode generator. A Moxtek 2500 Silicon drift energy sensitive detector was used in order to minimize the background level. The patterns were measured in the 2$\theta$ interval from 10 to 95$^\circ$, with a step size of 0.01$^\circ$ and 30 s counting time per step.

Rietveld refinements of the powder samples were performed with the program GSAS~\cite{Larson} using the EXPGUI graphic interface.~\cite{Toby}  Peak profiles were modeled using the pseudo Voigt function as implemented by Van Laar and Yelon,\cite{VanLaar} in which low angle peak asymmetry is calculated from the axial divergence as described by Finger \emph{et al.}\cite{Finger} The background was fit using an eight term Chebyschev polynomial of the first kind. A total of seven structural parameters were refined in the final cycle for the powder samples: the three oxygen atom coordinates and the Debye-Waller factor for every atom position.   No preferred orientation was observed and therefore none was refined.

\begin{figure}
\includegraphics[width=3.25in,keepaspectratio=true]{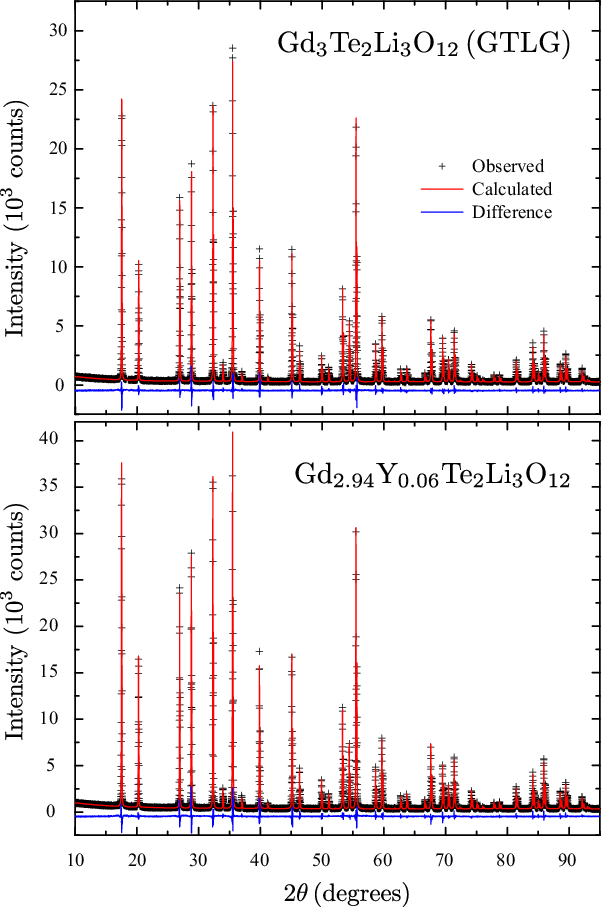}
\caption{Rietveld refinements of GTLG and Y-doped GTLG, showing experimental X-ray diffraction patterns, fit and residuals.
\label{XRDGTLG}}
\end{figure}

The GTLG sample showed a very clean XRD powder pattern with very sharp peaks (FWHM$ = 0.054^\circ$ at $2\theta = 35^\circ$) even at high angles (FWHM$ = 0.077^\circ$ at $2\theta = 85^\circ$). The sample is cubic with space group $\mathrm{Ia\bar{3}d}$ (230). A small peak was observed at $2\theta = 18.45^\circ$ ($d=4.81$ \AA), presumably from a small impurity. It could be due to a small percentage ($\leq 1$\%) of Li$_2$Te or LiGdO$_2$, both of which have their maximum intensity peak around $d=4.81$ \AA~\cite{ICSD}.  It was not possible to determine with certainty which compound makes up the impurity since its amount is very small and only one peak shows in the XRD pattern. The Gd$_{2.94}$Y$_{0.06}$Te$_2$Li$_3$O$_{12}$ or Y doped GTLG sample also showed a very clean XRD powder pattern, with very sharp peaks (FWHM$ = 0.056^\circ$ at $2\theta = 30^\circ$) even at high angles (FWHM$ = 0.084^\circ$ at $2\theta = 85^\circ$) and no impurity phase was observed. 

In the Rietveld refinement of  Gd$_{2.94}$Y$_{0.06}$Te$_2$Li$_3$O$_{12}$ an extra parameter was introduced as the Y and Gd shared occupancy of the 24$c$ site.  Both atoms were fixed to have the same Debye-Waller factor.  The plots from the final Rietveld fits are shown in Fig.~\ref{XRDGTLG}.  Both samples show an excellent agreement between the structural model and the experimental pattern with final Rietveld weighted profile $R$-factor, $R_{wp}$, values of 7.03\% for GTLG and 7.65\% for Y doped GTLG.

\begin{table*}
\caption{Atomic coordinates from the Rietveld refinement of the three powder garnet samples.  Only the oxygen sites have refined coordinate values. 
\label{AtomCoords}
}
\begin{ruledtabular}

\begin{tabular}{llllllll}
Compound & Atom & Site & $x$ (\AA) & $y$ (\AA) & $z$ (\AA) & Occupancy &  Thermal Displacement ($U_\mathrm{iso}$) \\
\hline

GTLG & Li  & 24$d$ & 0.37500 & 0.00000 & 0.25000 & 1.0 & 0.021(4) \\
	  & Te & 16$a$ & 0.00000 & 0.00000 & 0.00000 & 1.0 & 0.0177(2) \\
	  & Gd & 24$c$ & 0.12500 & 0.00000 & 0.25000 & 1.0 & 0.01923(17)\\
	  & O   & 96$h$ & 0.27892(18) & 0.10449(18) & 0.1985(2) & 1.0 & 0.0146(9)\\
\hline
Y:GTLG & Li  & 24$d$ & 0.37500 & 0.00000 & 0.25000 & 1.0 & 0.039(4) \\
	      & Te  & 16$a$ & 0.00000 & 0.00000 & 0.00000 & 1.0 & 0.0222(3) \\
	      & Gd  & 24$c$ & 0.12500 & 0.00000 & 0.25000 & 0.968(9) & 0.02342(18) \\
	      & Y  & 24$c$ & 0.12500 & 0.00000 & 0.25000 & 0.032(9) & 0.02342(18) \\
	      & O   & 96$h$ & 0.27892(18) & 0.10449(18) & 0.1985(2)  & 1.0 & 0.0208(9) \\
\hline
GAG & Al(1)  & 24$d$ & 0.37500 & 0.00000 & 0.25000 & 1.0 & 0.0240(5)\\
	& Al(2)  & 16$a$ & 0.00000 & 0.00000 & 0.00000 & 1.0 & 0.0224(6)\\
	& Gd     & 24$c$ & 0.12500 & 0.00000 & 0.25000 & 1.0 & 0.02343(14)\\
	& O  & 96$h$ & 0.28150(14) & 0.10178(15) & 0.20140(16)  & 1.0 & 0.0187(7)\\

\end{tabular}
\end{ruledtabular}
\end{table*}

The two compounds show very similar cell parameters: $a=12.3865(5)$ \AA\, for the pure GTLG samples and $a=12.3861(5)$ \AA\, for the Y doped material. Table~\ref{AtomCoords} shows the atomic coordinates for both the Gd$_3$Te$_2$Li$_3$O$_{12}$ and Gd$_{2.94}$Y$_{0.06}$Te$_2$Li$_3$O$_{12}$ samples respectively. They show that the structures of the two compounds are very similar, which is also confirmed by the inter-atomic distances. Table~\ref{Distances} shows the inter-atomic distances around the oxygen atom, which is the only atom with refined coordinate values.  All the distances are compatible with reported literature values. The Y content in the doped sample was 3.2(9)\% according to the Rietveld refinement, close to the expected value from the synthesis (2\%).

\begin{figure}
\begin{center}
\includegraphics[width=3.25in,keepaspectratio=true]{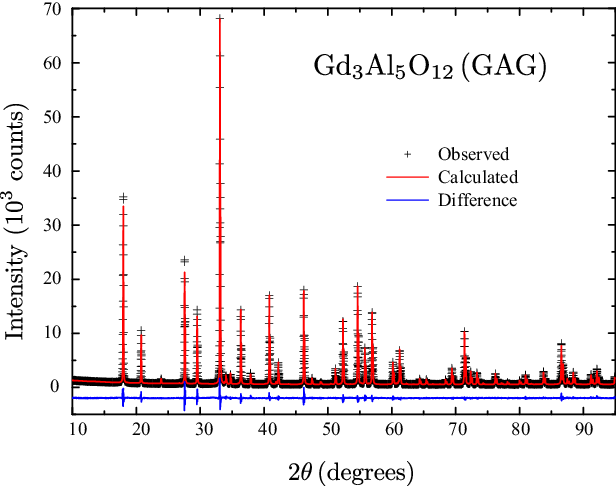}
\caption{Rietveld refinement of the Gd$_3$Al$_5$O$_{12}$ (GAG) sample, showing the measured X-ray diffraction pattern, the fits and residuals.
\label{XRDGAG}}
\end{center}
\end{figure}

The GAG sample also showed a very clean XRD powder pattern with very sharp peaks (FWHM$ = 0.064^\circ$ at $2\theta = 33^\circ$) even at high angles (FWHM$ = 0.0927^\circ$ at $2\theta = 86^\circ$). The sample has the expected cubic space group $\mathrm{Ia\bar{3}d}$ with lattice parameter of 12.1090(6) \AA. There is a small impurity of GdAlO$_3$ (ICSD code 59848~\cite{ICSD}).  The plots from the final Rietveld fits are shown in Fig.~\ref{XRDGAG}. Again, we see an excellent agreement between the structural model and the experimental pattern with final Rietveld $R_{wp}$ values of 5.95\%. Table~\ref{AtomCoords} shows the atomic coordinates obtained from the structure refinement. No evidence of atomic disorder between both Al positions and the Gd was observed in the Rietveld refinement.  In the refinement, a GdAlO$_3$ impurity was treated as a separate phase for which only the cell parameters and the scale factor were varied.  A quantitative analysis of the impurity phase using the Rietveld refinement gave a 2.4\% weight fraction for the GdAlO$_3$ impurity phase.

\begin{table}
\caption{Important inter-atomic distances for Gd$_3$Al$_5$O$_{12}$ and Gd$_3$Te$_2$Li$_3$O$_{12}$.
\label{Distances}
}
\begin{ruledtabular}

\begin{tabular}{lllll}
   & GTLG   &  Dirty GTLG &    & GAG     \\
Atoms & Distance (\AA) & Distance (\AA) & Atoms & Distance (\AA) \\
\hline
Li-O	& 1.870(3)  & 1.878(3)  &	Al(1)-O & 1.7741(17) \\
Te-O	 & 1.945(3) & 1.942(3) &	Al(2)-O & 1.9269(18) \\
Gd-O  & 2.391(3) &   2.391(3)  & Gd-O  & 2.3359(19) \\
Gd-O$'$ &  2.497(3) & 2.491(3)    &  Gd-O$'$  & 2.4843(19) \\
\end{tabular}
\end{ruledtabular}
\end{table}

While no clear indications of off-stoichiometry were detected in the analysis of the diffraction patterns, we cannot rule out small levels of substitutional disorder below the detectable levels of Rietveld refinement. The upper bound for excess Gd on Li or Al sites (in GTLG and GAG respectively) is 0.4\%, as calculated by the GSAS Rietveld refinement.  There was, however, insufficient contrast to rule out significant mixing between Gd and Te sites in GTLG.  Nonetheless, it seems unlikely to occur since the Gd$^{3+}$ and Te$^{6+}$ have very different valence charges.

\subsection{Single crystal GGG}

The GGG single crystal was oriented and polished perpendicular to the $[1,1,0]$ direction.  Diffuse scattering measurements were performed covering much of reciprocal space as high resolution diffraction peak profile measurements.  The measurements were performed using a Huber four circle goniometer sourced by a fine focus FR571 copper rotating anode coupled to OSMIC collimating multilayer optics. The high resolution peak profile measurements were performed with a double crystal Ge 220 monochromator added after the multilayer optics to provide a highly monochromatic beam. The diffuse scattering measurements were performed without the Ge monochromator in order to obtain good counting statistics.

\begin{figure}
\begin{center}
\includegraphics[width=3.25in,keepaspectratio=true]{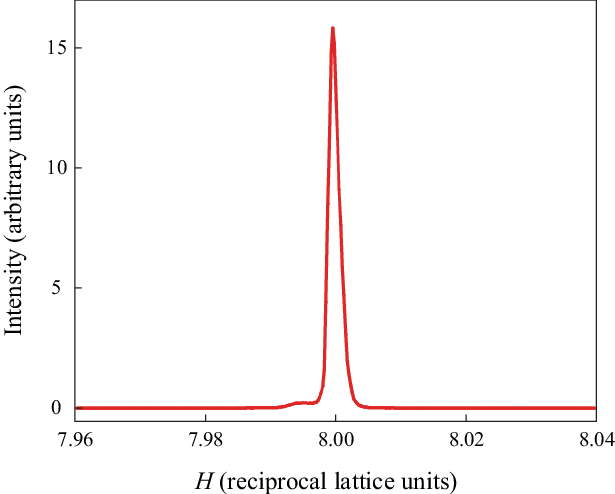}
\caption{Radial scan through the (8 4 0) reflection of GGG. \label{gggpeak} }
\end{center}
\end{figure}

No indication of diffuse scattering, which might have been an indication of chemical disorder or stacking faults in the sample, was detected. High resolution reciprocal space mapping around Bragg reflections showed very clean resolution-limited symmetric peaks with no appreciable effects due to off-stoichiometry domains in the sample. If such domains exist they should be larger than $\sim 150$ nm in size and exhibit lattice constants identical to the rest of the perfect crystalline GGG.  

The FWHM is very narrow, indicative of the extremely perfect crystal quality. Extensive reciprocal space mapping at a range of resolution configurations were not able to detect any measurable twinning, super structure peaks or diffuse scattering.  A slight shoulder on the low-Q side of the high resolution peak profile (see Fig.~\ref{gggpeak}) corresponds to a narrow, yet detectable, distribution of lattice constants most probably due to a distribution of stoichiometry (at the level of $\pm 0.0001$\% of Gd). The mosaic width of all reflections are resolution-limited at less than 0.01$^\circ$. In other words, the composition is extremely uniform and the quality of the crystal is exceptionally perfect. The width of the radial scan corresponds to correlation lengths larger than 0.8 $\mu$m.

In previous work,~\cite{Allibert1974,Darby2008} the lattice parameter of GGG samples has been correlated to the excess concentration of Gd, $x$, in the formula Gd$_{3+x}$Ga$_{5-x}$O$_{12}$, via the phenomenological expression
\begin{equation}
a = (12.375\,\text{\AA})\left[ 1 + \left(\frac{r_\mathrm{Gd}}{r_\mathrm{Ga}} - 1\right)0.0268x \right]
\end{equation}
where the ionic radii are given\cite{Shannon} as $r_\mathrm{Gd} = 1.053$ \AA\, and $r_\mathrm{Ga} = 0.62155$ \AA.  

In order to accurately determine the lattice constant, X-ray diffraction measurements following the Bond method~\cite{Bond} were performed on our single crystal of GGG.  Results from both the (8 4 0) and the (8 8 0) reflections were in excellent agreement and lead to a lattice constant of $12.3873\pm 0.0001$ \AA. This implies $x=0.053 \pm 0.005$ or an excess Gd concentration of $1.8\% \pm 0.1\%$. The diffractometer was aligned to within $\pm 0.01$ degree in both the meridional and axial directions of the incident X-ray beam by a method based on using Borrmann forward diffraction from a silicon crystal for a full range of $\chi$ angles.~\cite{Murphy}

To summarize this structural characterization of three Gd garnets, probable off-stoichiometry was found in GGG but not in GAG and GTLG.  The measured lattice parameter of GGG is indicative of 1.8\% excess Gd on Ga sites, although insufficient contrast between Gd and Ga inhibited direct observation of the randomness.  Disorder between Gd and Te in GTLG could also not be verified due to insufficient contrast.  There was enough contrast to rule out inter-site disorder, above the 0.4\% level, between Gd and Al in GAG and between Gd and Li in GTLG.


\begin{thebibliography}{10}

\bibitem{Moessner2001}
R.~Moessner,
\newblock Can. J. Phys {\bf 79}, 1283 (2001).

\bibitem{Villain1979}
J.~Villain,
\newblock Z. Phys. B {\bf 33}, 31 (1979).

\bibitem{Andreanov2010}
A.~Andreanov, J.~T. Chalker, T.~E. Saunders, and D.~Sherrington,
\newblock Phys. Rev. B {\bf 81}, 014406 (2010).

\bibitem{BellierCastella2001}
L.~{Bellier-Castella}, M.~J.~P. Gingras, P.~C. Holdsworth, and R.~Moessner,
\newblock J. Can. Phys. {\bf 79}, 1365 (2001).

\bibitem{Gingras1997}
M.~J.~P. Gingras, C.~V. Stager, N.~P. Raju, B.~D. Gaulin, and J.~E. Greedan,
\newblock Phys. Rev. Lett. {\bf 78}, 947 (1997).

\bibitem{Keren2001}
A.~Keren and J.~S. Gardner,
\newblock Phys. Rev. Lett. {\bf 87}, 177201 (2001).

\bibitem{Hyperkagome} The garnet structure consists of two interpenetrating lattices of corner sharing triangles known as hyperkagome lattices.  Although a different space group symmetry, the related hyperkagome lattice in the $S=1/2$ system Na$_4$Ir$_3$O$_8$ exhibits a possible quantum spin liquid ground state and has therefore attracted a great deal of recent interest. Y. Okamoto, M. Nohara, H. Aruga-Katori and H. Takagi, Phys. Rev. B {\bf 99}, 137207 (2007).

\bibitem{Applegate2007}
R.~P. Applegate, Y.~Zong, and L.~R. Corruccini,
\newblock J. Phys. Chem. Solids {\bf 68}, 1756 (2007).

\bibitem{Schiffer1995}
P.~Schiffer, A.~P. Ramirez, D.~A.Huse, P.~L. Gammel, U. Yaron, D.~J. Bishop, A.~J. Valentino
\newblock Phys. Rev. Lett. {\bf 74}, 2379 (1995).

\bibitem{Schiffer1994}
P.~Schiffer, A.~Ramirez, D.~Huse, and A.~Valentino,
\newblock Phys. Rev. Lett. {\bf 73}, 2500 (1994).

\bibitem{Tsui1999}
Y.~Tsui, C.~Burns, J.~Snyder, and P.~Schiffer,
\newblock Phys. Rev. Lett. {\bf 82}, 3532 (1999).

\bibitem{Tsui2001CJP}
Y.~Tsui, J.~Snyder, and P.~Schiffer,
\newblock Can. J. Phys {\bf 79}, 1439 (2001).

\bibitem{Daudin1982}
B.~Daudin, R.~Lagnier, and B.~Salce,
\newblock J. Magn. Magn. Mater. {\bf 27}, 315 (1982).

\bibitem{Petrenko1997}
O.~A. Petrenko, C.~Ritter, M.~Yethiraj, and D.~{McK. Paul},
\newblock Physica B {\bf 241}, 727 (1997).

\bibitem{Petrenko1998}
O.~A. Petrenko, C.~Ritter, M.~Yethiraj, and D.~{McK. Paul},
\newblock Phys. Rev. Lett. {\bf 80}, 4570 (1998).

\bibitem{Petrenko2009}
O.~A. Petrenko,  G.~Balakrishnan, D.~McK. Paul, M.~Yethiraj, G.~J. McIntyre and A.~S. Wills,
\newblock Journal of Physics: Conference Series {\bf 145}, 012026 (2009).

\bibitem{Costa2007}
A.~L. Costa, L.~Esposito, V.~Medri, and A.~Bellosi,
\newblock Adv. Eng. Mater. {\bf 2007}, 307 (2007).

\bibitem{Chiang2007}
C.-C. Chiang, M.-S. Tsai, and M.-H. Hon,
\newblock J. Electrochem. Soc. {\bf 154}, J326 (2007).

\bibitem{Darby2008}
M.~S.~B. Darby, T.~C. May-Smith, and R.~W. Eason,
\newblock Appl. Phys. A {\bf 93}, 477 (2008).

\bibitem{Allibert1974}
M.~Allibert, C.~Chatillon, J.~Mareschal, and F.~Lissalde,
\newblock J. Cryst. Growth {\bf 23}, 289 (1974).

\bibitem{Dunsiger2000GGG}
S.~Dunsiger,  J.~S. Gardner, J.~A. Chakhalian, A.~L. Cornelius, M.~Jaime, R.~F. Kiefl, R.~Movshovich, W.~A. MacFarlane, R.~I. Miller, J.~E. Sonier, and B.~D. Gaulin, 
\newblock Phys. Rev. Lett. {\bf 85}, 3504 (2000).

\bibitem{Crystallography}
W.~Borchardt-Ott,
\newblock {\em Crystallography} (Springer, Berlin, 1993).

\bibitem{LanthAct}
S.~Cotton,
\newblock {\em Lanthanide and Actinide Chemistry} (John Wiley and Sons, West
  Sussex, England, 2006).

\bibitem{Binder1986}
K.~Binder and A.~P. Young,
\newblock Reviews of Modern Physics {\bf 58}, 801 (1986).

\bibitem{Quilliam2007}
J.~A. Quilliam, C.~G.~A. Mugford, A.~Gomez, S.~W. Kycia, and J.~B. Kycia,
\newblock Phys. Rev. Lett. {\bf 98}, 037203 (2007).

\bibitem{Quilliam2007GSO}
J.~A. Quilliam, K.~A. Ross, A.~G. Del Maestro, M.~J.~P. Gingras, L.~R. Corruccini, and J.~B. Kycia,
\newblock Phys. Rev. Lett. {\bf 99}, 097201 (2007).

\bibitem{Yavorskii2006}
T.~Yavors'kii, M.~Enjalran, and M.~Gingras,
\newblock Phys. Rev. Lett. {\bf 97}, 267203 (2006).

\bibitem{Yavorskii2007}
T.~Yavors'kii, M.~Gingras, and M.~Enjalran,
\newblock J. Phys.: Condens. Matter {\bf 19}, 145274 (2007).

\bibitem{Overmeyer}
J. Overmeyer, \emph{Paramagnetic Resonance} (Academic Press, New York, 1963), chap. 15.

\bibitem{Wolf1962}
W.~P. Wolf, M.~Ball, M.~T. Hutchings, M.~J.~M. Leask, and A.~F.~G. Wyatt,
\newblock J. Phys. Soc. Japan {\bf 17 (Suppl. B-I)}, 443 (1962).

\bibitem{Kinney1979}
W.~I. Kinney and W.~P. Wolf,
\newblock J. Appl. Phys. {\bf 50}, 2115 (1979).

\bibitem{Marshall2002}
I.~M. Marshall, S.~J. Blundell, F.~L. Pratt, A.~Husmann, C.~A. Steer, A.~I. Coldea, W.~Hayes, and R.~C.~C. Ward,
\newblock J. Phys.: Condens. Matter {\bf 14}, L157 (2002).

\bibitem{Bonville2004GGG}
P.~Bonville, J.~A. Hodges, J.~P. Sanchez, and P.~Vulliet,
\newblock Phys. Rev. Lett. {\bf 92}, 167202 (2004).

\bibitem{Ghosh2008}
S.~Ghosh, T.~F. Rosenbaum, and G.~Aeppli,
\newblock Phys. Rev. Lett. {\bf 101}, 157205 (2008).

\bibitem{Deen2010}
P.~P. Deen, O.~A. Petrenko, G.~Balakrishnan, B.~D. Rainford, C.~Ritter, L.~Capogna, H.~Mutka, and T.~Fennell,
\newblock Phys. Rev. B {\bf 82}, 174408 (2010).

\bibitem{Maestro2007}
A.~{Del Maestro} and M.~J.~P. Gingras,
\newblock Phys. Rev. B {\bf 76}, 064418 (2007).

\bibitem{Onn1967}
D.~G. Onn, H.~Meyer, and J.~P. Remeika,
\newblock Phys. Rev. {\bf 156}, 663 (1967).

\bibitem{Landau1971}
D.~P. Landau, B.~E. Keen, B.~Schneider, and W.~P. Wolf,
\newblock Phys. Rev. B {\bf 3}, 2310 (1971).

\bibitem{Kamazawa2008}
K.~Kamazawa, D~Louca, R.~Morinaga, T.~J. Sato, Q.~Huang, J.~R.~D. Copley, and Y.~Qiu, 
\newblock Phys. Rev. B {\bf 78}, 064412 (2008).

\bibitem{Note2}
There are two notable exceptions to this rule: the garnet Yb$_3$Ga$_5$O$_{12}$
  and the pyrochlore Yb$_2$Ti$_2$O$_7$, which show sharp first-order phase
  transitions to spin liquid or hidden order ground states. 
  J.~A. Hodges, P.~Bonville, A.~Forget, A.~Yaouanc, P.~Dalmas de R\'{e}otier, G.~Andr\'{e}, M.~Rams, K.~Kr\'{o}las, C.~Ritter, P.~C.~M. Gubbens, C.~T. Kaiser, P.~J.~C. King, and C. Baines,
  \emph{et al.}, Phys. Rev. Lett. {\bf 88}, 077204 (2002); 
  P. Dalmas de R\'{e}otier, A.~Yaouanc, P.~C.~M. Gubbens, C.~T. Kaiser, C.~Baines, and P.~J.~C. King, Phys. Rev. Lett. {\bf 91}, 167201 (2003);  K.~A.~Ross, L.~R. Yaraskavitch, M.~Laver, J.~S. Gardner, J.~A. Quilliam, S.~Meng, J.~B. Kycia, D.~K. Singh, Th.~Proffen, H.~A. Dabkowska, and B.~D. Gaulin,
  Phys. Rev. B {\bf 84}, 174442 (2011).

\bibitem{Wills2006}
A.~S. Wills, M.~E. Zhitomirsky, B.~Canals, J.~P. Sanchez, P.~Bonville, P.~Dalmas de R\'{e}otier and A.~Yaouanc,
\newblock J. Phys.: Condens. Matter {\bf 18}, L37 (2006).


\bibitem{Note}
Glassy dynamics observed in the anisotropic kagom{\'e} antiferromagnet are the
  result of a different mechanism: S. Bekhechi and B. W. Southern, Phys. Rev. B
  {\bf 67}, 144403 (2003); P. Chandra and P. Coleman and I. Ritchey, J. Phys. I
  France {\bf 3}, 591 (1993);.

\bibitem{Tarjus2005}
G.~Tarjus, S.~A. Kivelson, Z.~Nussinov, and P.~Viot,
\newblock J. Phys.: Condens. Matter {\bf 17}, R1143 (2005).

\bibitem{Alba1999}
C.~Alba-Simionesco, J.~Fan, and C.~A. Angell,
\newblock J. Chem. Phys. {\bf 110}, 5262 (1999).

\bibitem{Larson}
A.~C. Larson and R.~B. {von Dreele},
\newblock Los Alamos National Laboratory Report Report No. LAUR 86-748, 1994
  (unpublished).

\bibitem{Toby}
B.~H. Toby,
\newblock J. Appl. Cryst. {\bf 34}, 210 (2001).

\bibitem{VanLaar}
B.~{Van Laar} and W.~B. Yelon,
\newblock J. Appl. Cryst. {\bf 17}, 47 (1984).

\bibitem{Finger}
L.~W. Finger, D.~E. Cox, and A.~P. Jephcoat,
\newblock J. Appl. Cryst. {\bf 27}, 892 (1994).

\bibitem{ICSD}
Inorganic Crystal Structure Database.

\bibitem{Shannon}
R.~D. Shannon,
\newblock Acta Cryst. {\bf A32}, 751 (1976).

\bibitem{Bond}
W.~L.~Bond, Acta Cryst. {\bf 13}, 814 (1960).

\bibitem{Murphy}
W.~J.~Murphy, \emph{et al.}, J. Appl. Cryst. {\bf 18}, 71 (1985).


\end{thebibliography}
\end{document}